\documentclass[12pt]{article}
\usepackage[margin=1.2in]{geometry}

\usepackage{graphicx,url}

\usepackage[utf8]{inputenc}  
\usepackage{authblk} % for authors' affiliations

\usepackage{amsmath,amsfonts,amssymb}
\usepackage{natbib}
\usepackage{hyperref}
\usepackage{bm} % to use \bm{} for bold greek letters
\usepackage{pdfpages}
\usepackage{array}
\usepackage{MnSymbol} % for \bigoplus, \bigominus

\newcommand{\bt}{\bold{\theta}}
\newcommand{\bT}{\bold{\Theta}}

\title{Model-based clustering for populations of networks}

\author[1]{Mirko Signorelli\footnote{m.signorelli@lumc.nl}}
\author[2,3]{Ernst C. Wit}

\affil[1]{Department of Biomedical Data Sciences, Leiden University Medical Center (NL)}
\affil[2]{Institute of Computational Science, University of Lugano (SW)}
\affil[3]{Bernoulli Institute for Mathematics, Computer Science and Artificial Intelligence,
      University of Groningen (NL)}

\date{}

\begin{document} 

\maketitle

\noindent \textbf{About this article}
\begin{itemize}
\item Please cite this article as: Signorelli, M. and Wit, E. C. (2020), Model-based clustering for populations of networks. \textit{Statistical Modelling}, 20 (1), 9-29. DOI: 10.1177/1471082x19871128.
\item This document contains the ``accepted'' version of the manuscript. The final (published) version of the article can be freely downloaded (Open Access) from the website of Statistical Modelling, using this link:\\ \href{https://doi.org/10.1177/1471082X19871128}{https://doi.org/10.1177/1471082X19871128}
\end{itemize}

\begin{abstract}
\noindent
Until recently obtaining data on populations of networks was typically rare. However, with the advancement of automatic monitoring devices and the growing social and scientific interest in networks, such data has become more widely available. From sociological experiments involving cognitive social structures to fMRI scans revealing large-scale brain networks of groups of patients, there is a growing awareness that we urgently need tools to analyse populations of networks and particularly to model the variation between networks due to covariates. We propose a model-based clustering method based on mixtures of generalized linear (mixed) models that can be employed to describe the joint distribution of a populations of networks in a parsimonious manner and to identify subpopulations of networks that share certain topological properties of interest (degree distribution, community structure, effect of covariates on the presence of an edge, etc.). Maximum likelihood estimation for the proposed model can be efficiently carried out with an implementation of the EM algorithm. We assess the performance of this method on simulated data and conclude with an example application on advice networks in a small business.\\

\end{abstract}

\section{Introduction}\label{sec:intro}

The last decades have witnessed a growing interest in the analysis of relational data. Typically, these data come in the form of a network that displays relations between individuals or objects, and they are represented by means of a graph wherein nodes, i.e., individuals or objects, are connected by edges, i.e., relations. In some applications, especially in genetics, relations cannot be observed directly and the main task is to infer them or their strength from the data \citep{friedman2008,abegaz2013,vujavcic2015computationally}. In this paper, we are interested in cases where relations between individuals or objects are observed and the networks themselves are the data.

For a long time, network science was almost exclusively concerned with the analysis of a single network, mainly because of the difficulty in collecting relational data and of limited computing capacity. Statistical modelling of  a single network \citep{snijders2011} has typically focused on certain aspects of network topology, such as degree distribution, network statistics or the presence of community structures. This has resulted into the development of a range of statistical network models that include the $p_1$ and $p_2$ models \citep{holland1981,vanduijn2004}, stochastic blockmodels \citep{holland1983,snijders1997}, exponential random graph models (ERGMs, \citealt{frank1986}), latent space models \citep{hoff2002} and the family of log-linear models proposed by \cite{perry2012}.

More recently, increased computing capacities, alongside with technological advances such as the development of sensor-based measurements, the diffusion of functional magnetic resonance imaging, the invention of high-throughput technologies in biology and the advent of social media, have multiplied the availability of relational data, spurring the analysis not only of larger networks, but also of several instances of the ``same'' network. The latter includes multilayer networks, dynamic networks and populations of networks. The availability of such collections of several networks poses new modelling challenges. Clearly, when data on several networks are available, modelling each network separately would be inefficient: irrespective of whether we are dealing with multilayer networks, longitudinal networks, or populations of networks, we expect networks therein to be similar to a certain degree; if this is indeed the case, analysing each network separately would not only be cumbersome, but also failing to use the statistical power of the ensemble. Instead, the specification of a joint model for the collection of networks makes it possible to achieve a more parsimonious representation of the data and to borrow information across networks in the estimation process; moreover, such a model may also be employed to identify groups of similar networks. Below, we briefly review some of the solutions that to date have been proposed to tackle this problem in the presence of dynamic networks, multilayer networks or populations of networks.

Dynamic networks allow to represent the evolution of a network system over time. \cite{snijders2001} proposed a stochastic actor-oriented model where the decision to create or dissolve an edge is based on some covariates, as well as on the current state of the network itself. \cite{hanneke2010} introduced a dynamic extension of ERGMs, known as Temporal Exponential Random Graph Model (TERGM). An extension of the Latent Space Models for dynamic networks has been proposed by \cite{sewell2015}. \cite{matias2017}, instead, developed a dynamic stochastic blockmodel that allows group membership of units to vary over time.

Multilayer networks are collections of networks that represent different types of relationships (multiple \textit{layers}) between a group of subjects. Two statistical models that allow to model jointly the layers of those networks are, among others, those of \cite{stanley2016} and \cite{paul2016}. \cite{stanley2016} proposed a multilayer stochastic blockmodel that assumes the existence of groups of networks, called strata, that share the same community structure. \cite{paul2016}, instead, introduced a multilayer stochastic blockmodel that assumes that the communities are the same in all layers, but allows different block-interaction probabilities in each layer.

Finally, a population of networks can be defined as a collection of independent graphs, each of which corresponds to a different statistical unit. Populations of networks arise, from example, when different individuals are asked to provide their view of relationships within a social network \citep{krackhardt1987} or when brain networks are compared across groups of patients \citep{taya2016}. Recently, there has been a growing interest in statistical modelling of populations of networks. \cite{sweet2014} proposed a hierarchical stochastic blockmodel that aims to infer groups of nodes that are shared across networks. Similarly, \cite{reyes2016} introduced a stochastic blockmodel for populations of networks that attempts to identify a unique community structure which is shared across networks. \cite{durante2017}, instead, extended the latent space model approach of \cite{hoff2002} to populations of networks by proposing a mixture model that describes the joint density of networks in the population using few components, each of which has a different latent-space representation. Finally, \cite{mukherjee2017} proposed to cluster graphs within a population of networks through a spectral clustering algorithm that is applied to a distance matrix that measures the distances between the graphon estimates of the graphs.

In this paper we focus on the problem of finding and characterizing clusters of graphs that are similar with respect to the effect of certain covariates of interest on the presence or absence of edges in a population of networks. Towards this aim, we propose to model the population of networks with a mixture model whose components can be any statistical network model that can be specified as a generalized linear model or a generalized linear mixed model; this includes, for example, the $p_1$ and $p_2$ models, degree-corrected stochastic blockmodels a priori, and the loglinear network models of \cite{perry2012}. The advantages of this methodological framework are that it makes it possible to describe populations of networks using some statistical models that are popular in social network analysis, that it can flexibly handle the inclusion of different types (monadic, dyadic and graph-specific) of covariates and that it furthermore allows to detect subpopulations of networks (if any). In Section \ref{sec:model-spec} we introduce and formalize our mixture of network models and we elaborate on the specification of the components of the mixture. Model estimation is considered in Section \ref{sec:model-est}, where we discuss how the proposed model can be estimated with an implementation of the Expectation-Maximization (EM) algorithm. In Section \ref{sec:sims} we assess the performance of our method on simulated data, and in Section \ref{sec:application} we present an example application to data on advice relationships in a small manufacturing firm described by \cite{krackhardt1987}.

\section{Model specification}\label{sec:model-spec}

We consider a sample of $K$ graphs $\{ \mathcal{G}_1, \mathcal{G}_2, \ldots, \mathcal{G}_K\}$, where each graph $\mathcal{G}_k = (V, E_k)$ comprises a set of edges $E_k$ between a set of $v$ vertices $V$, from a population of networks $\mathcal{S}$. We represent $\{ \mathcal{G}_1, \mathcal{G}_2, \ldots, \mathcal{G}_K\}$ with an array $\mathcal{Y}$ of dimension $v \times v \times K$, where each horizontal slice $\textbf{Y}_k$ is the adjacency matrix of graph $\mathcal{G}_k$. Therefore, an entry $Y_{ij}^k$ in $\mathcal{Y}$ refers to the presence (or intensity) or absence of edge $(i,j)$ in the $k$-th graph $\mathcal{G}_k$. If the graphs in $\mathcal{S}$ are undirected, each $\textbf{Y}_k$ is symmetric and we can restrict our attention to the upper triangle of $\textbf{Y}_k$.

In principle, one could imagine that each graph $\mathcal{G}_k$ is drawn from a different distribution $f(\textbf{Y}|\bt_k), \; k \in \{1,\ldots, K\}$ with parameter vector $\bt_k$: 
$$\textbf{Y}_k \sim f \left(\textbf{Y}|\bt_k\right).$$
In the presence of many networks, however, this would result in a cumbersome modelling exercise, yielding $K$ different models obtained from separate analyses of each graph. Since each graph is defined on the same set of vertices, it is natural to consider models with additional structure.

\subsection{Specification of the mixture model}

In this paper we consider the existence of clusters of graphs with similar $f \left(\textbf{Y}|\bt_k\right)$: if any such cluster exists, we would like to borrow information among graphs within that cluster, so as to estimate a joint model for graphs belonging to that cluster rather than many separate network models. As a result, we assume that the population of networks $\mathcal{S}$ arises from $M \leq K$ subpopulations $\mathcal{S}_1, \ldots, \mathcal{S}_M$ of graph models, each with probability density function $f \left(\textbf{Y}|\bt_m\right), \: m \in \{1,\ldots, M\}$. We denote by $Z_k \in \{ 1,\ldots,M \}$ the label that identifies the subpopulation of graph $\mathcal{G}_k$, such that $Z_k = m$ if $\textbf{Y}_k \sim f \left(\textbf{Y}|\bt_m\right)$ (i.e., $Z_k = m$ if $\mathcal{G}_k \in \mathcal{S}_m$). Since in real problems it will typically be unknown which graph belongs to which subpopulation, the vector of identifying labels $\textbf{Z} = (Z_1,\ldots,Z_K)$ is a latent variable. Therefore, we view each graph in the sequence as a random draw from a mixture model whose components are the probability density functions $f \left(\textbf{Y}|\bt_m\right)$:
\begin{equation}
\textbf{Y}_k \sim \sum_{m=1}^M \pi_m f \left(\textbf{Y}| \bt_m \right), \label{graphmix}
\end{equation}
with mixing proportions $\pi_m = Pr(Z_k = m)$ denoting the prior probabilities that a graph belongs to the $m$-th subpopulation $\mathcal{S}_m$. Clearly, we assume $\pi_m \geq0$ $\forall m \in \{1,\ldots,M\}$ and $\sum_{m=1}^M \pi_m = 1$.
If we let $\bT = (\bt_1,\ldots,\bt_M)$, the likelihood of model \eqref{graphmix} is thus
\begin{equation}
\begin{aligned}
L(\bT |\mathcal{Y},\textbf{Z}) = Pr(\mathcal{Y}, \textbf{Z} | \bT) &= \prod_{k=1}^K Pr(\textbf{Y}_k|Z_k,\bT) Pr(Z_k |\bT)\\ &= \prod_{k=1}^K \pi_{Z_k} f \left(\textbf{Y}_k| \bt_{Z_k} \right). 
\end{aligned}
\label{likelihood}
	\end{equation}

This likelihood suffers from the usual identifiability issues when considering mixture models. Each of the $K$ components can be permuted without altering the likelihood. So, as there are $K!$ possible permutations, there exists $K!$ symmetries in the likelihood. Moreover, the possibility of empty components raises the possibility that certain parameters $\bt_k$ are not identifiable. As our aim is to find the maximum likelihood estimate, we will be satisfied with finding one of the $K!$ equivalent MLEs. The issue of empty (or near-empty) components is dealt with via information criteria to select the number of components. Although not providing any theoretical guarantees, (near) empty components will be discouraged due to the unnecessary numbers of parameters they introduce.  

\subsection{Specification of the components of the mixture}\label{sub:spefic-comp}

The way in which the probability density functions $f \left(\textbf{Y}| \bt_m \right)$ in Equations \eqref{graphmix} and \eqref{likelihood} can be specified depends on the properties that are deemed relevant for the analysis of the networks at hand. 
If, for example, interest lies in clustering a sequence of binary graphs based on similarities in their degree distributions, $f \left(\textbf{Y}| \bt_m \right)$ can be specified as a $p_1$ or a $p_2$ model \citep{holland1981,vanduijn2004}. If a partition of vertices into groups or communities is available and the probabilities of interaction between vertices are believed to depend on group memberships, a stochastic blockmodel \citep{holland1983} can be employed to specify $f$. If both the degree distribution and community structure are deemed relevant, different types of degree-corrected stochastic blockmodels \citep{wang1987,signorelli2017} can be considered. If one would like to cluster graphs based on the values of network statistics that reflect socially relevant patterns of interaction (for example, transitivity), they could consider exponential random graphs (ERGMs, \citealt{frank1986}).

In this paper we focus our attention on network models that assume edges to be independent conditionally on the model parameters (and, potentially, on a set of unobserved random effects), so that their likelihood can be specified as that of a generalized linear (mixed) model. The motivation behind this choice is three-fold. Firstly, a wide range of popular network models (among which are the $p_1$ and $p_2$ models, stochastic blockmodels a priori, degree-corrected stochastic blockmodels a priori, the family of models considered by \cite{perry2012} and the unconstrained model that we introduce in Section \ref{sub:unc}, but not ERGMs) can be specified as generalized linear models (GLMs, \citealt{mccullagh1989}) or as generalized linear mixed models (GLMMs, \citealt{breslow1993}). Moreover, the GLM(M) framework enables us to easily incorporate monadic, dyadic and graph-specific covariates into the network generative models. Finally, mixtures of GLM(M)s can be estimated efficiently and this ensures computational efficiency in the estimation of mixtures of network models, which we will base on an iterative algorithm that may require several iterations and, thus, could otherwise become computationally burdensome. 

Therefore, we shall specify the mixture model in \eqref{graphmix} as a mixture of GLMs \citep{grun2008} by assuming that the value of each edge $y_{ij}^k$ is drawn from an exponential family distribution and that a transformation of the conditional expectation of $Y_{ij}^k$ is linear in the parameters:
\begin{equation*}
g \left[ E \left( Y_{ij}^k|\textbf{x}_{ijk}, \bt, Z_k = m \right) \right] = \textbf{x}_{ijk}^T \bt_m,
\end{equation*}
where $g$ is a link function and $\textbf{x}_{ijk}$ is a vector associated to $\bt_m$ that can contain \emph{monadic} (i.e., node-specific) covariates for nodes $i$ and $j$, \emph{dyadic} (i.e., edge-specific) covariates for edge $(i,j)$ and graph-specific covariates for graph $\mathcal{G}_k$. Extensions to mixtures of GLMMs are straightforward and will be used in the application. 
The density of graph $\mathcal{G}_k$ can then be obtained as
$f \left(\textbf{Y}_k| \bt_{z_k} \right) = \prod_{i<j} f \left(y_{ij}^k| \bt_{z_k} \right)$ if $\mathcal{G}_k$ is undirected, or as $f \left(\textbf{Y}_k| \bt_{z_k} \right) = \prod_{i \neq j} f \left(y_{ij}^k| \bt_{z_k} \right)$ if it is directed. Below we shortly introduce three network models that we will use in Section \ref{sec:sims} to illustrate our method.

\subsubsection{$p_1$ model}\label{sub:p1}

In social network analysis, the popularity of individuals is often regarded as one of the possible determinants of the formation of relations between individuals in a network. This reflects the idea that in certain social settings, individuals may be more likely to relate to popular individuals than to isolated ones: for example, if you live in a small village in the heart of the Alps, you are more likely to interact with popular figures such as the mayor and the priest, rather than with a woodsman who lives in a remote cottage in the middle of the woods.
This idea is at the basis of the $p_1$ model \citep{holland1981}, a simple network model that assumes that the probability of an edge between any two nodes $i$ and $j$ depends (only) on the expected degrees of the two nodes. If, for example, a population of binary undirected networks is considered, we can specify a mixture of $p_1$ models by letting $ y_{ij}^k | z_k \sim \text{Bern} \left( \pi_{ij}^{z_k} \right) $, where 
$\text{logit} \left( \pi_{ij}^{z_k} \right) = \theta^{z_k} + \alpha_{i}^{z_k} + \alpha_{j}^{z_k}$ and $\sum_{i=1}^v \alpha_{i}^{z_k} = 0$.

\subsubsection{Stochastic blockmodel}\label{sub:sbm}

Besides popularity, group membership of nodes is another factor that can shape the way in which relations are formed. Real networks often feature the presence of \textit{communities} of nodes whose members are highly connected with each other and tend to form sporadic connections with members from other communities. For example, it has been shown that Parliamentarians tend to collaborate more frequently with members from their same parliamentary group, rather than with those from other political groups \citep{signorelli2018}. In general, group membership typically induces a so-called \textit{community structure} in networks, wherein nodes from the same community are closely tied to each other and sporadically linked to nodes from other communities.
The effect of community membership on the formation of relations is usually modelled with stochastic blockmodels \citep{holland1983,snijders1997}. Let $\mathcal{P}$ denote a partition of $V$ into $p < v$ groups and denote by $C: V \to \mathcal{P}$ a community-assignment function, so that $C(i)$ is the community that node $i$ belongs to. In stochastic blockmodels, the probability of an edge between nodes $i$ and $j$ depends on the communities that the two nodes belong to: $ y_{ij}^k | z_k \sim \text{Bern} \left( \pi_{ij}^{z_k} \right) $, where 
\begin{equation}
\text{logit} \left( \pi_{ij}^{z_k} \right) = \theta^{z_k}_{C(i)C(j)}.
\label{form:sbm}
\end{equation}
Depending on whether the community-assignment function is known or not, it is possible to distinguish stochastic blockmodels \textit{a priori} \citep{holland1983}, wherein community labels are known and interest lies in the reconstruction of relationships between communities, from stochastic blockmodels \textit{a posteriori} \citep{snijders1997}. In this work we focus on the simpler a priori stochastic blockmodel, which is computationally cheap and, thus, can be easily incorporated into the iterative estimation procedure proposed in Section \ref{sub:EM}. Mixtures of stochastic blockmodels a posteriori, instead, are considered in the works of \cite{stanley2016} and \cite{reyes2016}.

\subsubsection{Unconstrained network model}\label{sub:unc} 

The $p_1$ model and stochastic blockmodel described above are two examples of simple and thrifty statistical network models that can employed to model commonly observed features of real networks such as heterogeneity in node degrees and community structure. These models comprise a number of parameters that is considerably lower than the number of nodes pairs and, thus, they allow a very parsimonious description of networks; however, in reality these models are likely to be often too simplistic. It may thus be desirable to consider more complex statistical models, which can improve model fit and enable a more realistic description of the complex structure of a network. For example, it is possible to combine the aforementioned models into a degree-corrected stochastic blockmodel \citep{wang1987} that can account for degree heterogeneity and community structure at the same time, or to incorporate covariates into stochastic blockmodels \citep{signorelli2018}. A further example of how to combine different statistical network models into a more realistic one can be found in the example application that we provide in Section \ref{sec:application}, where we will specify a network model that combines features of the $p_2$ model and of the stochastic blockmodel, and that furthermore accounts for the effect of some monadic covariates on the formation of advice relationships.

Clearly, more realistic network models may require a larger set of parameters and this could increase the complexity of maximum likelihood estimation for model \eqref{graphmix} and computing time. To illustrate this, we consider the extreme scenario of a mixture of saturated network models, where the number of parameters is equal to the number of edge pairs multiplied by the number of subpopulations of graphs, namely $M v(v-1)/2$ in undirected graphs and $M  v(v-1)$ in directed graphs. This model simply assumes that $y_{ij}^k | z_k \sim \text{Bern} \left( \pi_{ij}^{z_k} \right)$, leaving the probabilities $\pi_{ij}^{z_k}$ unconstrained. It represents the most complex model that can be specified to model relations within a population of networks with $M \leq K$ subpopulations, and it does not make any restrictive assumption about which factors affect the creation of edges. As such, in practice this model may represent a useful starting point in the analysis of the population of networks: in particular, its generality can be exploited at an initial stage of the analysis to choose the number of subpopulations $M$ in the mixture and to identify some important patterns in the data; information gathered from this complex model could then be exploited to further refine the analysis by specifying a simpler network model that accounts for the most important effects that are believed to affect the presence of edges. We provide an example of this modelling approach in Section \ref{sec:application}.

\section{Model estimation}\label{sec:model-est}

We propose to estimate the unknown parameter vector $\bT$ of the mixtures of network models described in Section \ref{sec:model-spec} with maximum likelihood. Since the likelihood function $L(\bT |\mathcal{Y},\textbf{Z})$ in equation \eqref{likelihood} depends both on the observed graphs $\mathcal{Y}$ and on the unobserved vector $\textbf{Z}$, such likelihood can be maximized by implementing the EM algorithm as illustrated below.

\subsection{EM algorithm}\label{sub:EM}

The Expectation-Maximization (EM) algorithm \citep{dempster1977} represents a popular choice for the estimation of mixture models. The algorithm allows the maximization of a likelihood $L(\bt | \textbf{y},\textbf{z})$ in the presence of latent labels $\textbf{z}$, and it consists of successive iterations of two steps, respectively called the expectation (E) and maximization (M) steps. The expectation step requires the computation of the conditional expectation of the likelihood $L(\bt | \textbf{y},\textbf{z})$ given the current estimate of $\bt$ and the observed data $\textbf{y}$, whereas the maximization step updates the parameter estimates by maximizing the expected likelihood determined in the E step. We propose the following implementation of the EM algorithm for the maximization of \eqref{likelihood}:
\begin{enumerate}
	\item choose a starting point for the algorithm made by the initial probabilities $p_{km}^0 = Pr(Z_k = m) \in [0,1]$ for $k \in \{1,\ldots,K\}$ and $m \in \{1,\ldots,M\}$, with $\sum_{m=1}^M p_{km}^0 = 1 \: \forall k$. Denote by $\textbf{P}^0$ the $K \times M$ matrix which collects these probabilities;
	\item given $\textbf{P}^0$, estimate the parameters of the mixture of GLMs with weights given by $\left(p_{1m}^0,\ldots,p_{Km}^0 \right)$ for the $m$-th component, and obtain $\hat{\bT}^0 = \left( \hat{\bt}^0_1,\; \ldots,\; \hat{\bt}^0_M \right)$;
	\item for $t = 1, 2, 3, \ldots$ until convergence is reached:
	\begin{itemize}
	\item[$\square$] \textbf{E step}. Given $\hat{\bT}^{t-1}$, derive $\textbf{P}^{t}$ as
	\begin{equation*}
p^{t}_{km} = \frac{f(\textbf{Y}_k|\hat{\bt}_m^{t-1})}{\sum_{j=1}^M f(\textbf{Y}_k|\hat{\bt}_j^{t-1})}.
	\end{equation*}	
	\item[$\square$] \textbf{M step}. Given $\textbf{P}^t$, estimate a mixture of GLMs with weights given by $\left(p_{1m}^t,\ldots,p_{Km}^t \right)$ for the $m$-th component, and obtain $\hat{\bT}^t$.
	\end{itemize}
\end{enumerate}

In principle it is possible to inizialize the EM algorithm introduced above with any matrix of initial probabilities $\textbf{P}^0$. However, it is possible to reduce the number of iterations and facilitate convergence to the true MLE by considering multiple sensible initial guess of the cluster memberships. Therefore, we consider three different cluster initializations by means of three network similarity measures combined with  the Partition Around Medoids (PAM) clustering method \citep{reynolds2006}. The first similarity measure is the Jaccard index \citep{jaccard1912distribution}. The second is given by the $L^1$ distance between the adjacency matrices (note that for binary graphs, this is equivalent to the $L^2$ distance). The third similarity measure is obtained by first computing the Laplacian matrix of each graph (defined as $L = D - A$, where $A$ is the adjacency matrix of the graph and $D$ a diagonal matrix with the degrees of each node as diagonal entries), and then taking the $L^1$ distance between the Laplacian matrices rather than between the adjacency matrices. Once a distance matrix has been obtained with one of the aforementioned methods, we apply the PAM clustering algorithm with number of clusters equal to $M$ and derive $\textbf{P}^0$ accordingly.

\subsection{Selection of the number of components} \label{sub:inf-crit}

In practice, the number of subpopulations $M$ that form the mixture is typically unknown and it needs to be estimated. The estimation of the number of components $M$ can be performed by minimizing model selection criteria such as the Akaike (AIC) and the Bayesian (BIC) Information Criteria. The choice of the effective sample size for the use inside the BIC is particularly crucial for multivariate data \citep{berger2014effective} and it is selected to be equal to $K$ for this purpose. We assess the performance of AIC and BIC on simulated data in Section \ref{sub:sim-inf-crit}.

\section{Simulations}\label{sec:sims}

In this section we first evaluate the accuracy of the proposed clustering method with respect to network size (represented by the number of nodes $v$), to the number of networks $K$ and to the number of subpopulations $M$ on simulated data. Then, we assess the capacity of the selection criteria introduced in Section \ref{sub:inf-crit} to correctly identify the true number of subpopulations $M$. We conclude discussing the scalability of the proposed method to large populations of networks and to populations of large networks. The \texttt{R code} to simulate the data and to perform model-based clustering of populations of networks can be found at \url{http://www.statmod.org/smij/archive.html}.

\subsection{Clustering accuracy}\label{sub:purity}

We begin the assessment of the performance of the proposed method with 9 simulations (A-I) where we study the clustering accuracy of our method with respect to the three network models introduced in Section \ref{sub:spefic-comp} as $v$, $K$ or $M$ increases. We focus on how the \textit{purity} \citep{schutze2008introduction} of the clusters is affected by these parameters. Purity is a measure of of clustering accuracy that attains value 1 if perfect classification is achieved; for $M$ equally sized (true) subpopulations, the worst-case purity value is $1/M$.

Table \ref{Tab:sim-overview} summarizes the features of the mixtures of networks considered in each simulation. Within each simulation, we compute 50 repetitions for each combination of $(v, K, M)$ considered; we consider 10 different initializations for the EM, 3 of which are obtained as described in Section \ref{sub:EM} and the remaining 7 are obtained from the previous 3 starting points by randomly replacing the initial probabilities of $30\%$ of the graphs. A more detailed description of the parameters involved in each simulation can be found in Section 1 of the Supplementary Materials.

\begin{table}
	\centering
		\begin{tabular}{|c|c|c|c|c|}\hline
Simulation & Network model & $v$ & $K$ & $M$\\\hline
A & $p_1$ model & from 10 to 40 & 50 & 2\\
B & $p_1$ model & 20 & from 12 to 60 & 2\\
C & $p_1$ model & 30 & $10 \cdot M$ & from 2 to 7\\
D & SBM a priori & from 10 to 40 & 50 & 2\\
E & SBM a priori & 20 & from 12 to 60 & 2\\
F & SBM a priori & 30 & $10 \cdot M$ & from 2 to 7\\
G & unconstrained & from 10 to 40 & 50 & 2\\
H & unconstrained & 15 & from 12 to 60 & 2\\
I & unconstrained & 15 & $10 \cdot M$ & from 2 to 7\\\hline
		\end{tabular}
\caption{Synthetic overview of simulations A-I. We consider two parsimonious models, the $p_1$ model (Section \ref{sub:p1}) and the stochastic blockmodel (SBM) a priori (Section \ref{sub:sbm}), and a more general, unconstrained network model (Section \ref{sub:unc}) that contains as many parameters as edge pairs. In simulations $A$, $D$ and $G$ we increase $v$, keeping $K$ and $M$ fixed. In simulations $B$, $E$ and $H$ we increase $K$, keeping $v$ and $M$ fixed. In simulations $C$, $F$ and $I$ we increase the number of subpopulations $M$ while keeping $v$ fixed; each subpopulation consists of 10 graphs (hence, $K = 10 \cdot M$).
}\label{Tab:sim-overview}
\end{table}

\begin{figure}
\begin{center}
\begin{tabular}{ccc}
	\includegraphics[scale=0.28, page = 1]{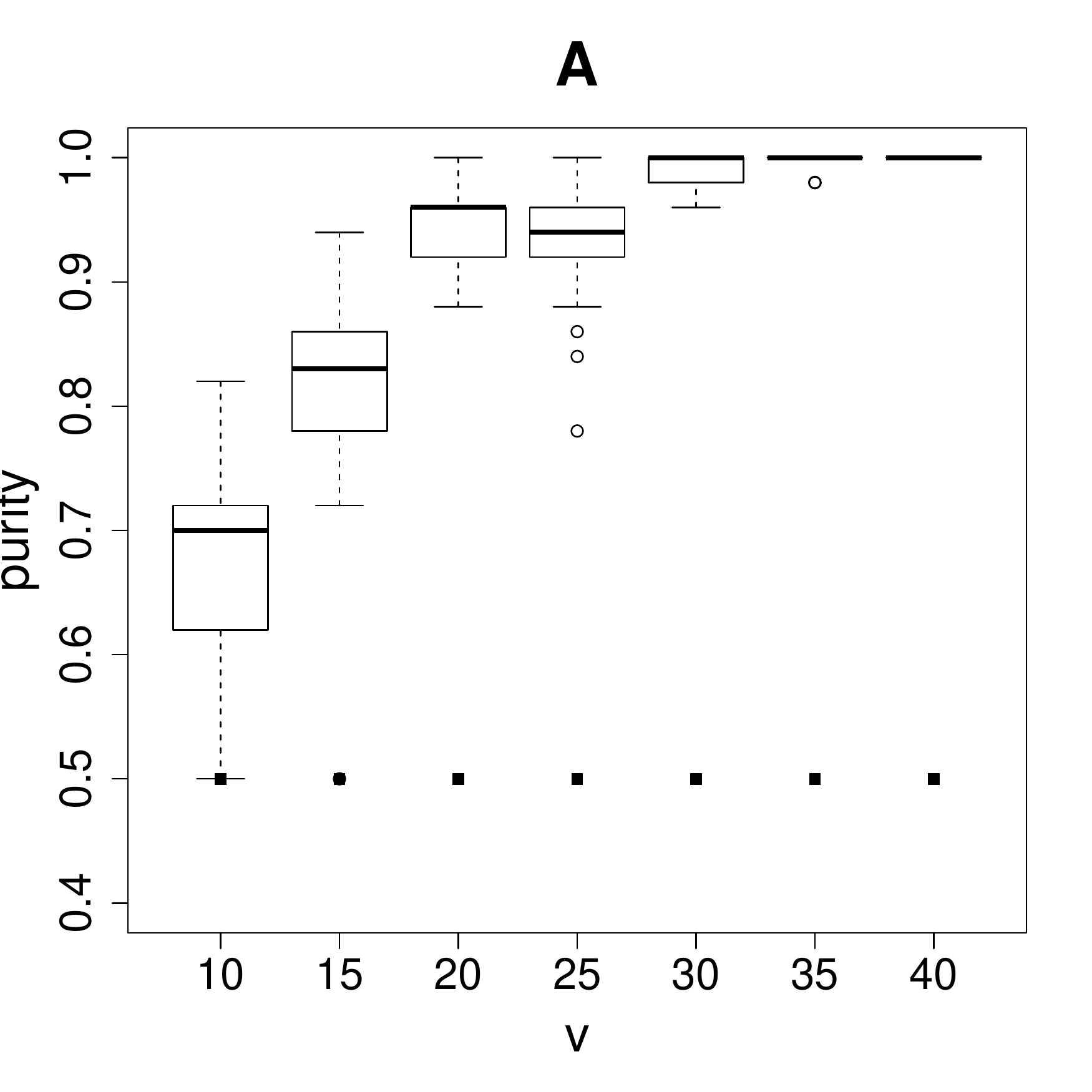} &
	\includegraphics[scale=0.28, page = 1]{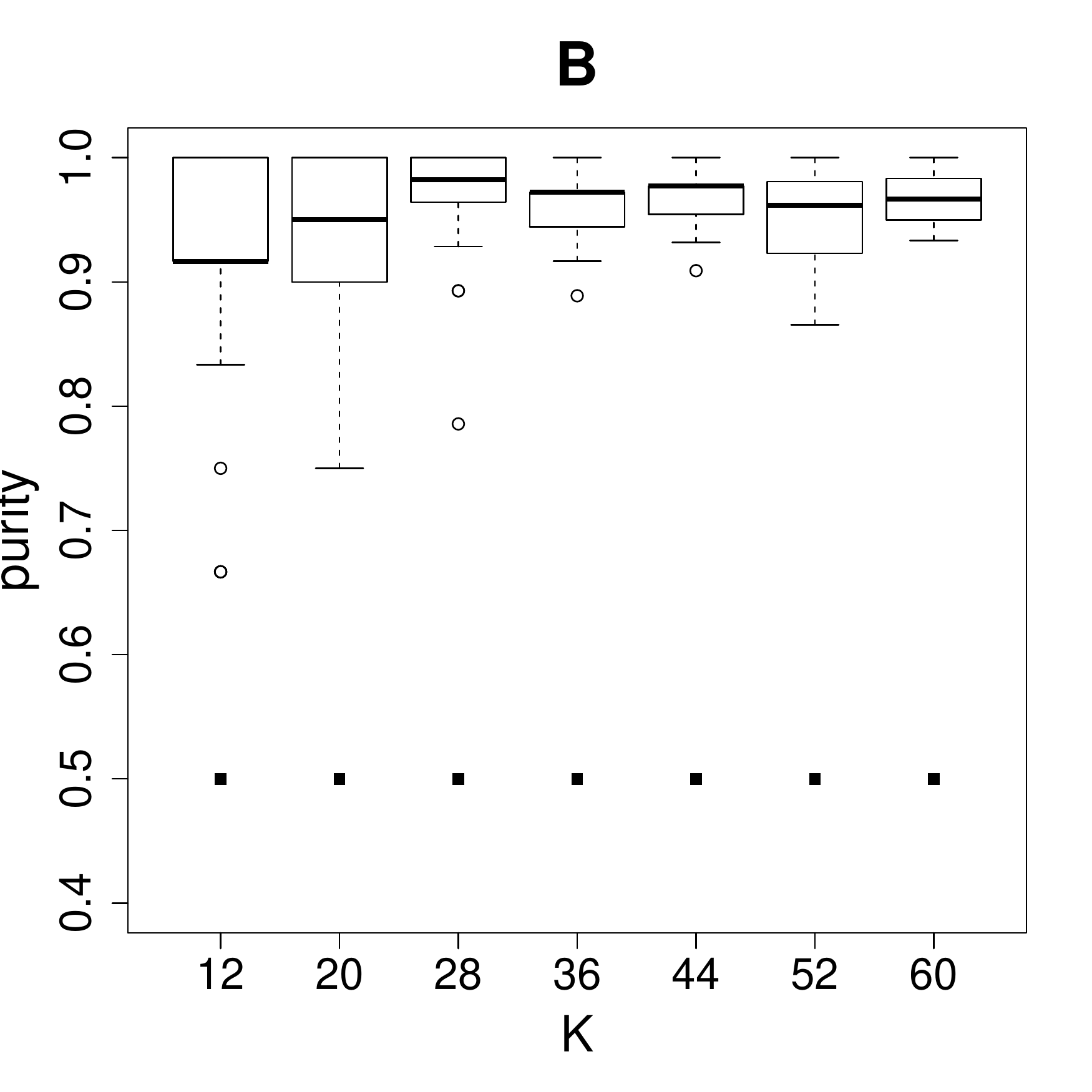} &
	\includegraphics[scale=0.28, page = 1]{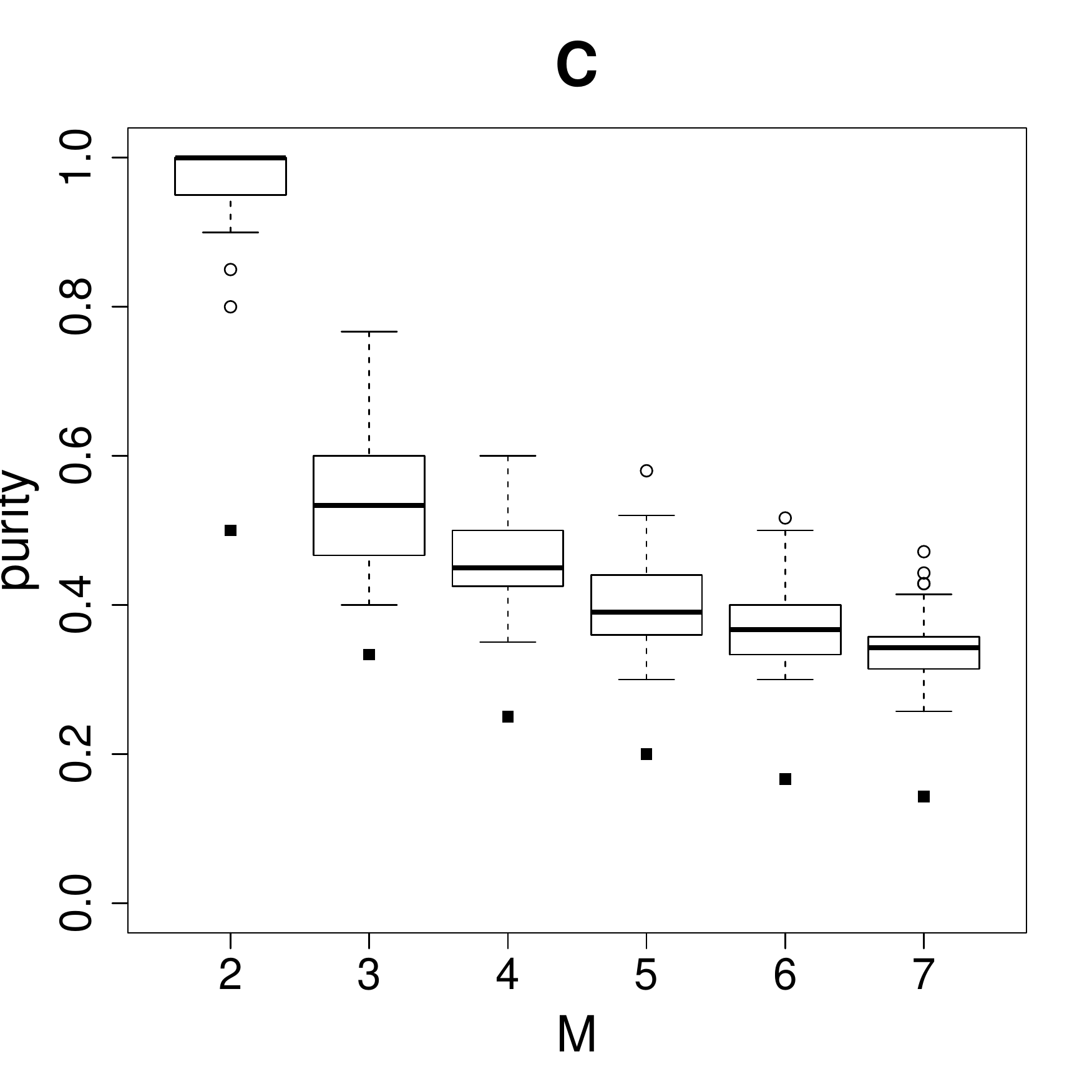} 
\end{tabular}
\end{center}
	\caption{Purity in simulations A, B, C. Each boxplot represents the distribution of purity over 50 repetitions, whereas the squares denote the value of purity that corresponds to a random assignment of graphs to clusters (i.e., $1/M$).}\label{fig:simABC}
\end{figure}

The distribution of purity across repetitions for the $p_1$ model is illustrated in Figure \ref{fig:simABC}. Purity quickly increases with respect to the number of nodes present in a graph (panel A); this steep increase is mainly due to the fact that the number of edge pairs increases quadratically with $v$, making prediction for populations of larger graphs a much easier task. Panel B shows that purity is already fairly high with a small number of graphs, but is highly variable; a larger $K$ results into reduced variability for the purity, which is more concentrated around its median value. Finally, simulation C shows that purity decreases with the number of subpopulations considered; this result is intuitive, since a larger number of subpopulations produces a harder classification problem; nevertheless, even for large $M$ there is an evident improvement over random allocation of graphs to subpopulations (panel C).

Similar observations hold for the simulations (D, E and F) with the stochastic blockmodel a priori (Supplementary Figure 1) and for those (G, H and I) with the unconstrained network model (Supplementary Figure 2).

\subsection{Selection of the number of subpopulations}\label{sub:sim-inf-crit}

In order to assess the performance of the model selection criteria introduced in Section \ref{sub:inf-crit}, in simulation J we repeatedly sample $K$ networks from a mixture of unconstrained network models (defined in Section \ref{sub:unc}) with $M = 3$ subpopulations of equal size (more details can be found in Section 1 of the Supplementary Materials). We fix $v = 20$ and let $K \in \{30, 90, 180, 300\}$. We repeat each simulation 100 times, computing the maximum likelihood estimates of the mixture model parameters for $M \in \{1, 2, 3, 4, 5\}$. Then, we compute AIC and BIC and derive the optimal number of subpopulations according to each criterion.

\begin{figure}
\begin{center}
\includegraphics[scale=0.35, page = 1]{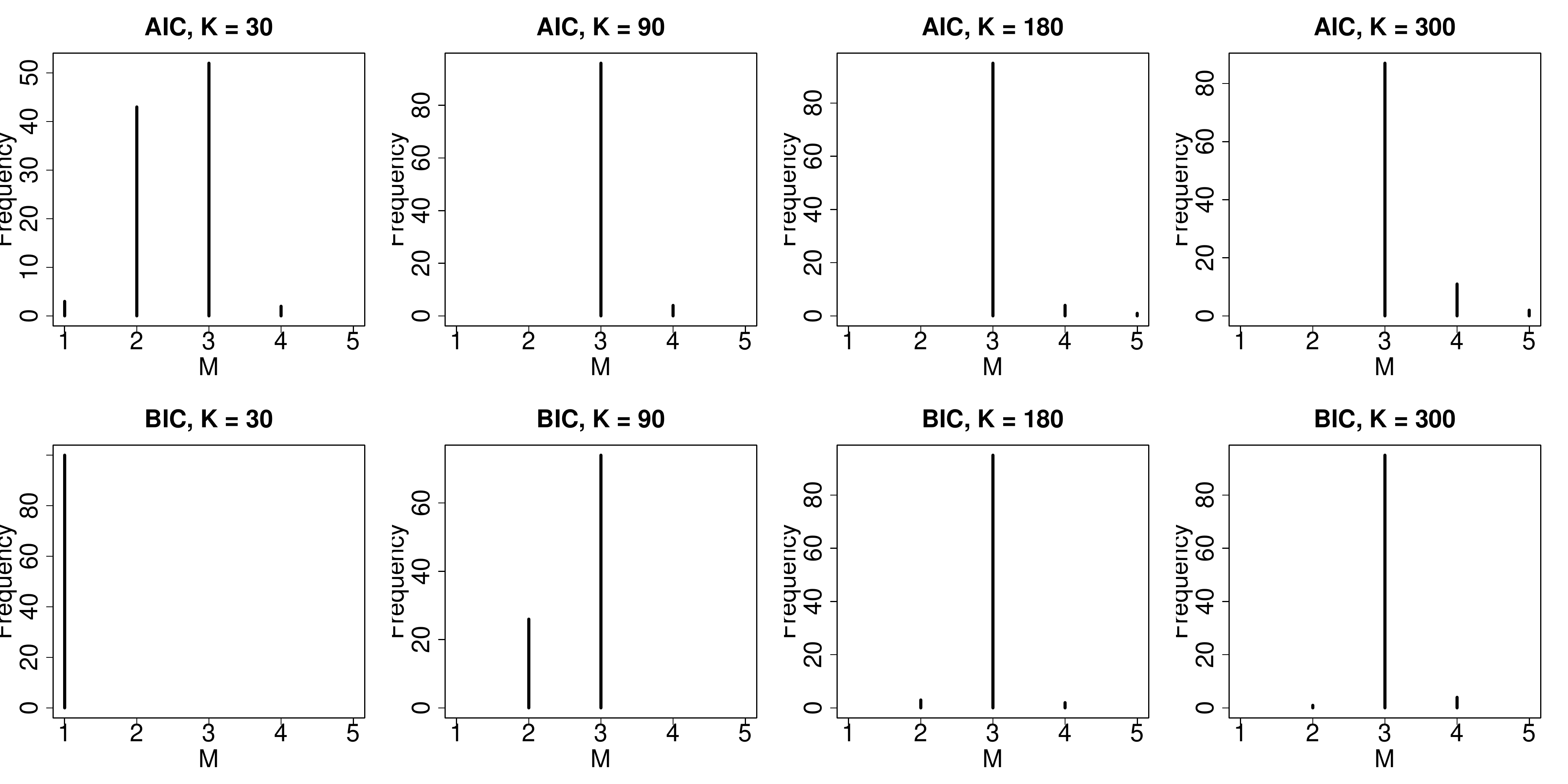}
\end{center}
	\caption{Distribution of the optimal number of subpopulations in simulation J according to Akaike's (AIC) and the Bayesian (BIC) information criteria, for different values of $K$. The true number of populations is equal to 3.}\label{fig:simJ}
\end{figure}

Figure \ref{fig:simJ} shows the distribution of the optimal number of subpopulations obtained with AIC and BIC for the different values of $K$ considered.
We note that as expected, both AIC and BIC can accurately select the correct number of subpopulations ($M = 3$) when a sufficiently large number of graphs $K$ is available. When $K$ is small, however, BIC tends to systematically underestimate $M$ and it is thus outperformed by AIC.

\subsection{Scalability of method to large and many graphs}\label{sub:scalability}

In Section \ref{sub:purity} we have considered simulation scenarios with a relatively small number of graphs of moderate size. This has allowed us to show how the proposed approach can achieve a good accuracy in allocating graphs to their correct subpopulation already in problems where $v$ or $K$ are relatively small. Here, we consider two simulations with larger $v$ and $K$ to illustrate the scalability of our approach, focusing on how computing time is affected by the number of networks $K$ as well as by the size of the networks. In general, we see that computing time increases linearly with $K$ and $M$, and super-linearly with $v$.

\begin{figure}
\begin{center}
\includegraphics[scale=0.4, page = 1]{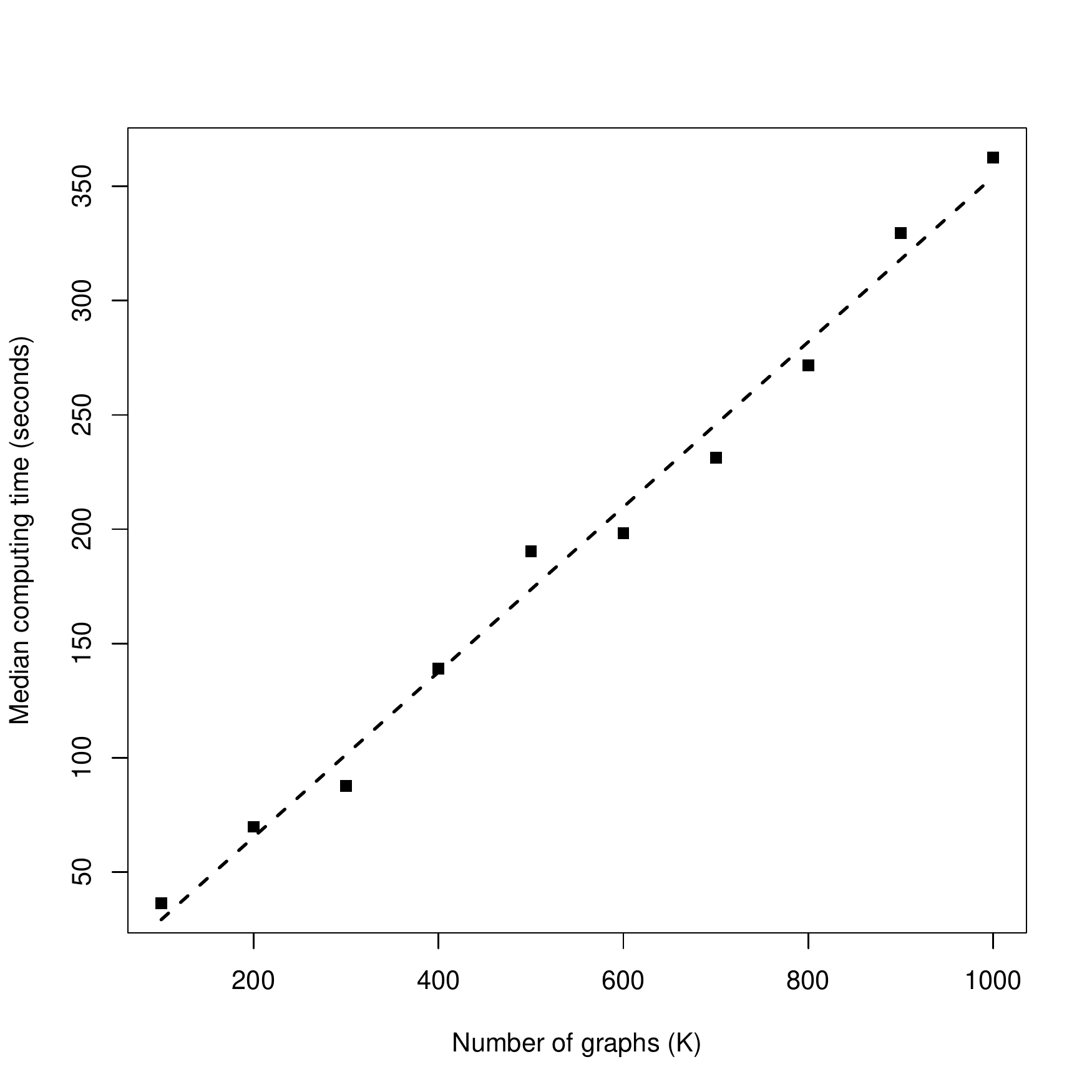}
\end{center}
	\caption{Relationship between number of graphs and median computing time in simulation K. It can be observed that computing time is approximately linear in the number of graphs. Computations were performed using a processor with 2.3 GhZ CPU.}\label{fig:simK}
\end{figure}

In simulation K we simulate data from a mixture of stochastic blockmodels a priori with 5 blocks setting $v = 50$, $M = 2$, and we let $K$ increase from 100 to 1000. Figure \ref{fig:simK} shows that the median computing time is linear in the number of graphs, and it increases from 36.5 seconds when $K = 100$ up to 363 seconds when $K = 1000$.

\begin{figure}
\begin{center}
\includegraphics[scale=0.35, page = 1]{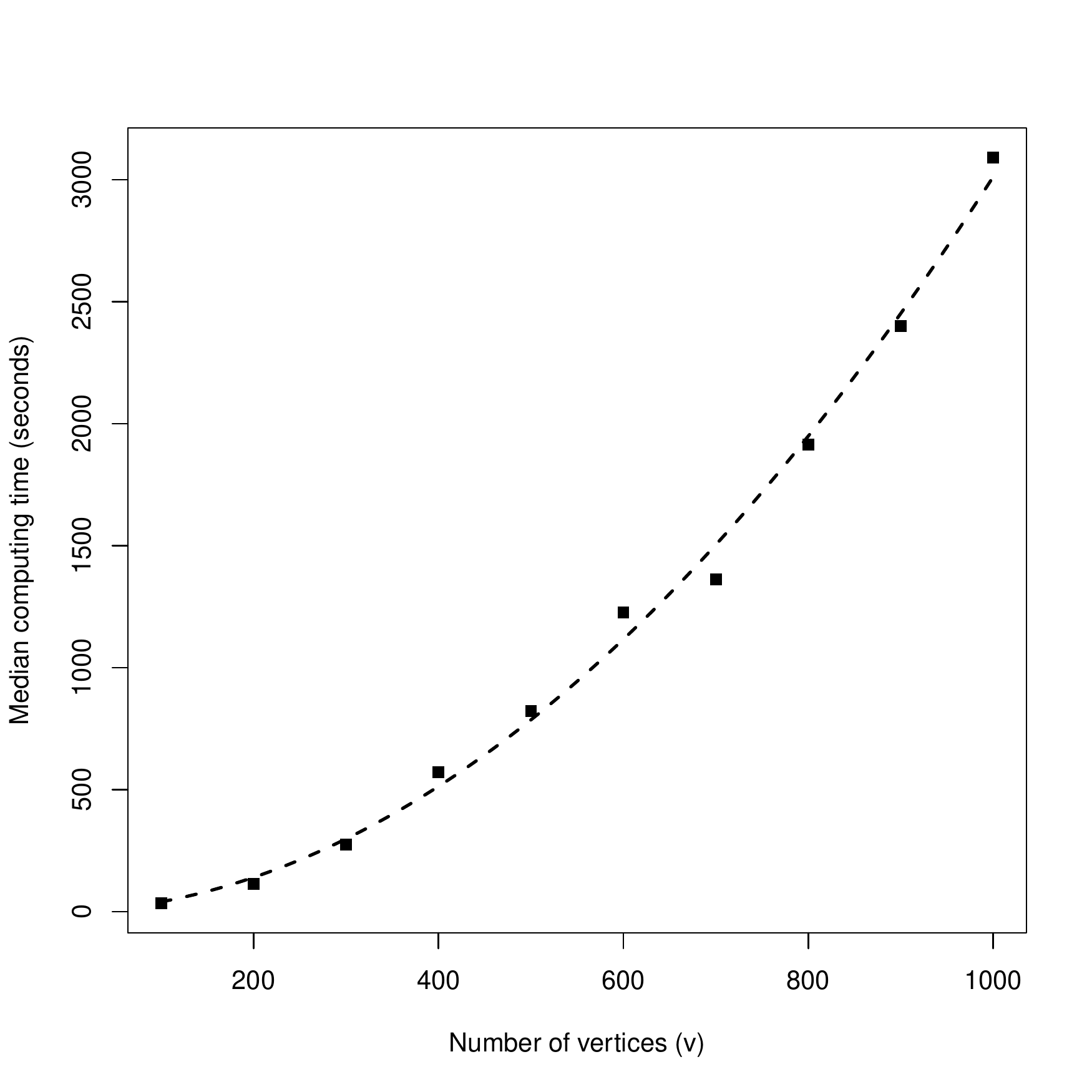}
\includegraphics[scale=0.35, page = 2]{fig4-simL.pdf}
\end{center}
	\caption{Relationship between graph size and median computing time in simulation L. It can be observed that computing time is approximately quadratic in the number of vertices $v$, and linear in the number of edge pairs $v(v-1)/2$. Computations were performed using a processor with 2.3 GhZ CPU.}\label{fig:simL}
\end{figure}

In simulation L we simulate data from a mixture of stochastic blockmodels a priori with 5 blocks setting $K = 50$, $M = 2$, and we let $v$ increase from 100 to 1000. Figure \ref{fig:simL} shows that the median computing time is quadratic in the number of vertices $v$ and linear in the number of node pairs $v(v-1)/2$, increasing from 35.1 seconds when $v = 100$ to 3089 seconds when $v = 1000$.

\section{Application}\label{sec:application}

In this Section we illustrate an application of our model-based network clustering method to a population of networks on advice relationships within a small business collected by \cite{krackhardt1987}, whose aim was to find ways to summarize the reconstructions of an unobserved social network reported by different perceivers. In this study, $21$ employees of a high-tech US company were asked to fill in a questionnaire where, among other questions, each employee was requested to reconstruct advice relationships between the 21 employees. From the answers to this questionnaire, \cite{krackhardt1987} obtained $K = 21$ directed advice networks, wherein each network is the reconstruction of advice relationships according to a different employee. Given the difficulty to analyse data within the resulting 3 dimensional array, which he called \textit{cognitive social structure}, \cite{krackhardt1987} proposed three simple aggregation techniques to reduce the dimensionality of the problem and simplify interpretation. Alternatively, we show here how a suitably defined mixture of network models may be employed to highlight important patterns in Krackhardt's population of networks.

Because each employee attempted to reconstruct the actual network of advice relationships within the firm, which is unobserved, we may expect that not only different perceivers would reconstruct advice relationships in a different manner, but also that some perceivers may provide substantially similar reconstructions of the advice network. In other words, it seems reasonable to hypothesize the presence of different clusters of networks, corresponding to groups of employees who have similar perceptions of advice relationships within the firm.

We begin our analysis in an exploratory manner by considering at first a mixture of $M$ unconstrained network models: we do not make any assumption on how each arrow is formed, thus leaving the probabilities to observe an arrow from node $i$ to node $j$ ($i \neq j$) unconstrained. The aim of this first analysis is two-fold: first, we want to find an appropriate number of clusters, unconfounded by a too restrictive network model; secondly, we aim to exploit patterns in the estimated probabilities of observing an arrow in each subpopulation to further refine the analysis. Later in this section we will use this information to define a more parsimonious network model where we will let these probabilities depend on a set of covariates.

\begin{figure}
\begin{center}
\includegraphics[scale=0.4]{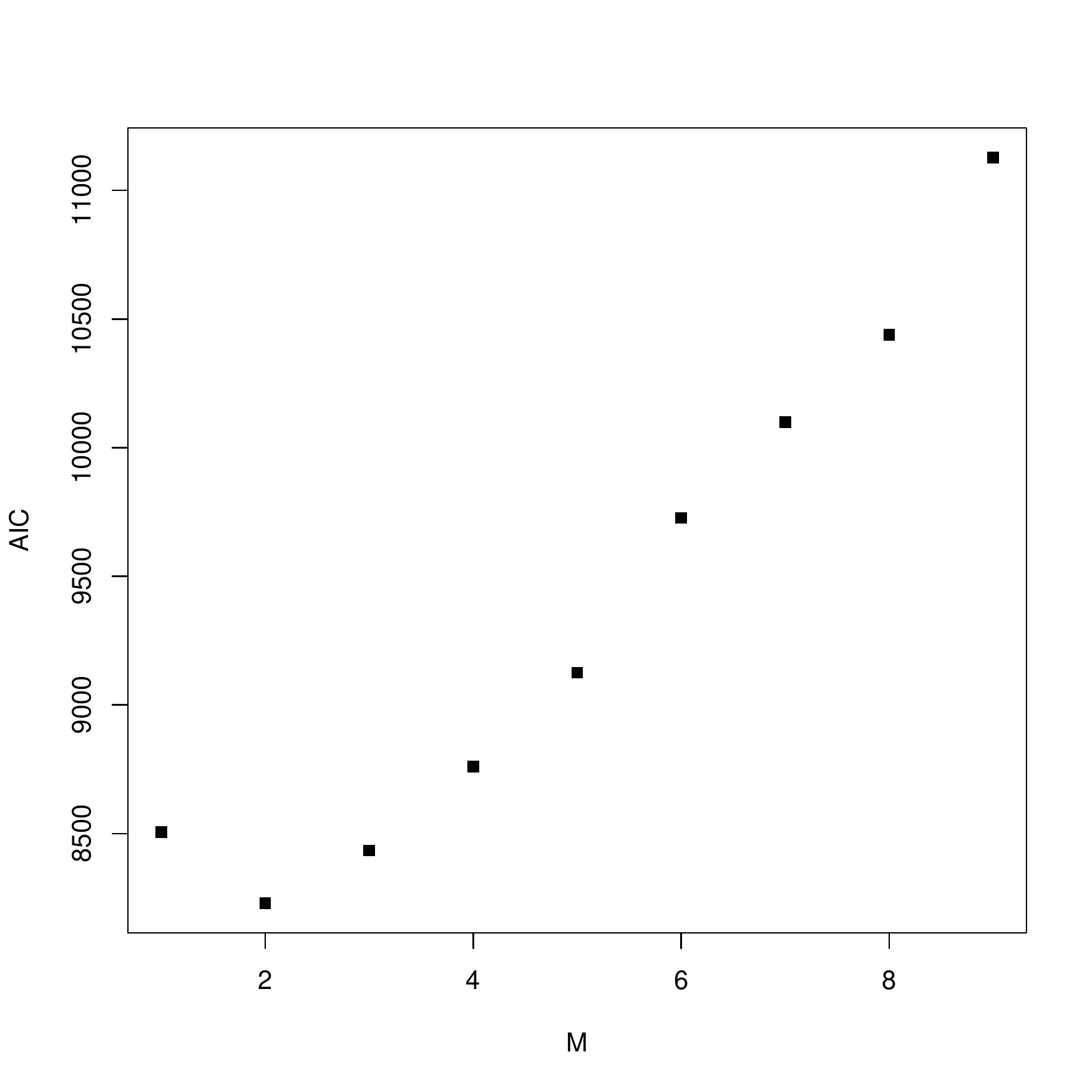} 
\end{center}
	\caption{Value of the Akaike Information Criterion for mixtures of unconstrained network models with increasing number of subpopulations $M$. The minimum AIC is attained when $M = 2$.}
\label{fig:appl-aic}
\end{figure}

\begin{figure}
\begin{center}
\includegraphics[scale=0.425, page = 1]{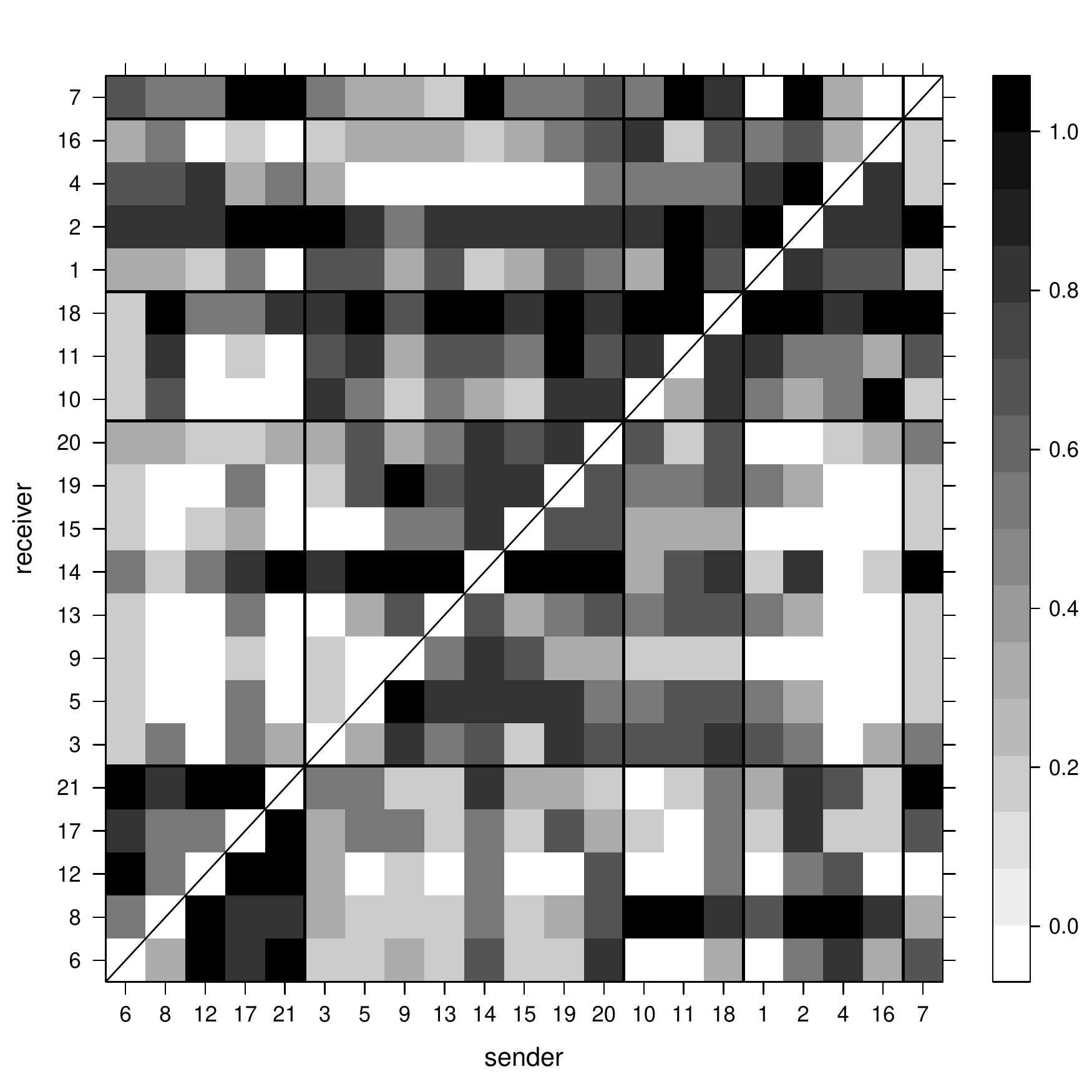} 
\includegraphics[scale=0.425, page = 2]{fig6-appl-phat-by-dept.pdf} \end{center}
\caption{Predicted probabilities to observe an arrow from individual $i$ ($x$ axis) to individual $j$ ($y$ axis) in the two subpopulations. On both axes, nodes are ordered by department and horizontal and vertical lines separate employees into the 4 departments the firm is divided into (so, for example, employees 6, 8, 12, 17 and 21 belong to department 1; note that employee 7, the CEO, doesn't belong to any department). It is apparent that graphs within the first subpopulation are denser, and that in both subpopulations department affiliation induces a community structure wherein advice relationships are typically more frequent within the same department than between different departments.} 
\label{fig:appl-unc}
\end{figure}

The first model that we consider simply assumes that $y_{ij}^k | z_k \sim Bern(\pi_{ij}^{z_k})$, with $z_k \in \{1, \ldots, M\}$ and $i \neq j$. We estimate the optimal number of subpopulations $\hat{M}$ following the approach outlined in Section \ref{sub:inf-crit}, using AIC as model selection criterion based on the results presented in Section \ref{sub:sim-inf-crit}. As Figure \ref{fig:appl-aic} shows, AIC attains a minimum when $M =2$, so we set $\hat{M} = 2$. Estimation of the mixture model with $M = 2$ components leads to the detection of a first cluster that comprises six perceivers, namely employees 1, 3, 4, 5, 10 and 21, and of a further cluster comprising the other 15 employees. In Figure \ref{fig:appl-unc} we show the predicted probabilities of observing advice relationships from sender $i$ to receiver $j$ in each subpopulation, i.e., $\hat{\pi}_{ij}^m$, with $m \in \{1, 2\}$ and $i \neq j$.

Graphical inspection of Figure \ref{fig:appl-unc} clearly reveals that graphs in the first subpopulation are denser than graphs in the second subpopulation; moreover, in both subpopulations we can intuitively observe that department affiliation seems to have a strong influence on the predicted probabilities of advice relationships. However, the large number of parameters (840) employed by the mixture of unconstrained models makes it difficult to draw any further conclusion on similarities and differences between the two subpopulations, and to relate those to any other known feature of the employees. Therefore, we now consider a more parsimonious model where we try to relate the presence of an arrow to such features. \cite{krackhardt1987} collected the following additional information about the employees:
\begin{itemize}
\item age and length of service (tenure) of each employee;
\item position occupied by each employee in the firm; one employee is the CEO, two are vice-presidents and the remaining 18 have supervision roles; here, we consider a binary distinction between CEO and vice-presidents on the one hand, and the other 18 employees on the other;
\item the department that each employee belonged to; in total, the firm comprises 4 departments.
\end{itemize}

We incorporate these covariates into the analysis by considering a network model where we combine features of the $p_2$ model \citep{vanduijn2004}, of the stochastic blockmodel a priori \citep{holland1983} and we furthermore let arrows depend on the available set of monadic covariates \citep{signorelli2018}. Such a model can be seen as a degree-corrected stochastic blockmodel a priori with covariates, where the blocks are given by the four departments in the company, the (in- and out-) degree-correction is carried out using random effects and where we furthermore account for the effect of several monadic covariates.  Let $A_i$ and $T_i$ denote the age and tenure of node $i$, let $L_i$ be a binary variable distinguishing individuals in leadership positions that is 1 if $i$ is either the CEO or a vice-president and 0 otherwise, and let $I(i = k)$ and $I(j = k)$ be binary variables that are 1 if, respectively, the perceiver ($k$) is sender ($i$) or receiver ($j$), and 0 otherwise. Furthermore, let $D_i \in \{1, 2, 3, 4\}$ denote the department that individual $i$ is affiliated to. We consider the following mixture model: $y_{ij}^k | \left(z_k, u_i^{z_k}, v_i^{z_k} \right) \sim Bern(\pi_{ij}^{z_k})$, where
\begin{equation}
\begin{gathered}
\text{logit}(\pi_{ij}^{z_k}) = \beta_0^{z_k} + u_i^{z_k} + v_j^{z_k} +
\beta_1^{z_k} A_i + \beta_2^{z_k} T_i + \beta_3^{z_k} L_i + \beta_4^{z_k} I(i = k)\\
+\beta_5^{z_k} A_j + \beta_6^{z_k} T_j + \beta_7^{z_k} L_j + \beta_8^{z_k} I(j = k)\\
+ \sum_{r = 1}^4 \gamma_r^{z_k} I[D_i = r] + \sum_{s = 1}^4 \delta_s^{z_k} I[D_j = s] + \sum_{r = 1}^4 \sum_{s = 1}^4 \xi_{rs}^{z_k} I[D_i = r] I[D_j = s],
\end{gathered}
\label{form:model-appl}
\end{equation}

\vspace{-0.2cm}
$u_i^{z_k} \sim N \left[ 0, (\sigma^{z_k})^2 \right]$ and $v_j^{z_k} \sim N \left[ 0, (\tau^{z_k})^2 \right]$ are random intercepts that allow to model parsimoniously the in- and out-degree distributions, and $\gamma_r^m$, $\delta_s^m$ and $\xi_{rs}^m$ are blockmodel main effects and interactions subject to the constraints that $\sum_{r=1}^4 \gamma_r^{m} = 0$, $\sum_{s=1}^4 \delta_s^{m} = 0$ and $\sum_{r=1}^4 \sum_{s=1}^4 \xi_{rs}^{m} = 0$ for every $m \in \{1, 2\}$.

We remark that not only model \eqref{form:model-appl} is considerably thriftier than the unconstrained mixture model previously considered (the former comprises 54 parameters, the latter 840), but it is also more interpretable as it enables us to study directly the relationship between the advice relationships reconstructed by the employees and individual (age and tenure) and organizational (department and leading roles) features of the employees and firm.

\begin{table}[h]
	\centering
\begin{tabular}{|c|c|c|c|c|c|}\hline
Parameter & $\hat{\theta}^1$ & $\hat{\theta}^2$ & $SE(\hat{\theta}^1)$ & $SE(\hat{\theta}^2)$ & p-value ($\theta^1 = \theta^2$)\\
 \hline
$\beta_0$ & 0.809 & -1.997$^*$ & 0.671 & 0.439 & 0.000 \\ 
$\beta_1$ (age sender) & -0.014 & -0.006 & 0.012 & 0.010 & 0.972 \\ 
$\beta_2$ (tenure sender) & -0.035$^*$ & -0.016 & 0.017 & 0.009 & 0.930 \\ 
$\beta_3$ (sender in lead pos.) & 0.014 & 0.008 & 0.016 & 0.013 & 0.977 \\ 
$\beta_4$ (perceiver = sender) & 1.128$^*$ & 1.020$^*$ & 0.231 & 0.146 & 0.675 \\ 
$\beta_5$ (age receiver) & 0.034 & 0.044$^*$ & 0.022 & 0.012 & 0.964 \\ 
$\beta_6$ (tenure receiver) & 0.543$^*$ & 0.582$^*$ & 0.214 & 0.170 & 0.876 \\ 
$\beta_7$ (receiver in lead pos.) & 1.407$^*$ & 2.058$^*$ & 0.287 & 0.150 & 0.017 \\ 
$\beta_8$ (perceiver = receiver) & 1.353$^*$ & 1.354$^*$ & 0.231 & 0.149 & 0.998 \\ \hline
$\sigma$ (rand. int. sender) & 0.329 & 0.267 &  &  &  \\ 
$\tau$ (rand. int. receiver) & 0.472 & 0.232 &  &  &  \\ 
\hline
		\end{tabular}
\caption{Maximum likelihood estimates and standard errors for $\beta^{m}, \: m \in \{1,2\}$ in model \eqref{form:model-appl}, and maximum likelihood estimates of $\sigma^{m}$ and $\tau^{m}$. Asterisks ($^*$) denote regression coefficients that are significantly different from 0 ($H_0: \beta_j^m = 0$) at $\alpha = 5\%$ level. The last column contains the p-value of the test for equality of each parameter in the two subpopulations ($H_0: \: \beta_j^1 = \beta_j^2$).}
\label{tab:p2sbm-estimates}
\end{table}

Table \ref{tab:p2sbm-estimates} contains the maximum likelihood estimates of the fixed effects $\beta_0^{m}, \ldots, \beta_8^{m}, \; m \in \{1, 2\}$ and of the standard deviation of the random effects in model \eqref{form:model-appl}. In both subpopulations we observe that the perceiver tends to report more ingoing and outgoing relationships that involve him ($\hat{\beta}_4 > 0$ and $\hat{\beta}_8 > 0$). Moreover, there is a common tendency to seek advice from employees with longer tenure within the firm ($\hat{\beta}_6 > 0$) and from the CEO and the vice-presidents ($\hat{\beta}_7 > 0$). As concerns differences between the two subpopulations, not only it is apparent that graphs in the first subpopulation are significantly denser than those in the second ($\hat{\beta}_0^1 > \hat{\beta}_0^2$), but we also observe that in the second subpopulation the tendency to seek advice from the CEO and vice-presidents is significantly stronger than in the first one ($\hat{\beta}_7^2 > \hat{\beta}_7^1$). Furthermore, $\hat{\tau}^1 > \hat{\tau}^2$  indicates a less heterogeneous distribution of outdegrees in the second subpopulation (i.e., in subpopulation 1 advice requests tend to be more concentrated on fewer employees).

\begin{figure}%[h]
\begin{center}
\includegraphics[scale=0.425, page = 1]{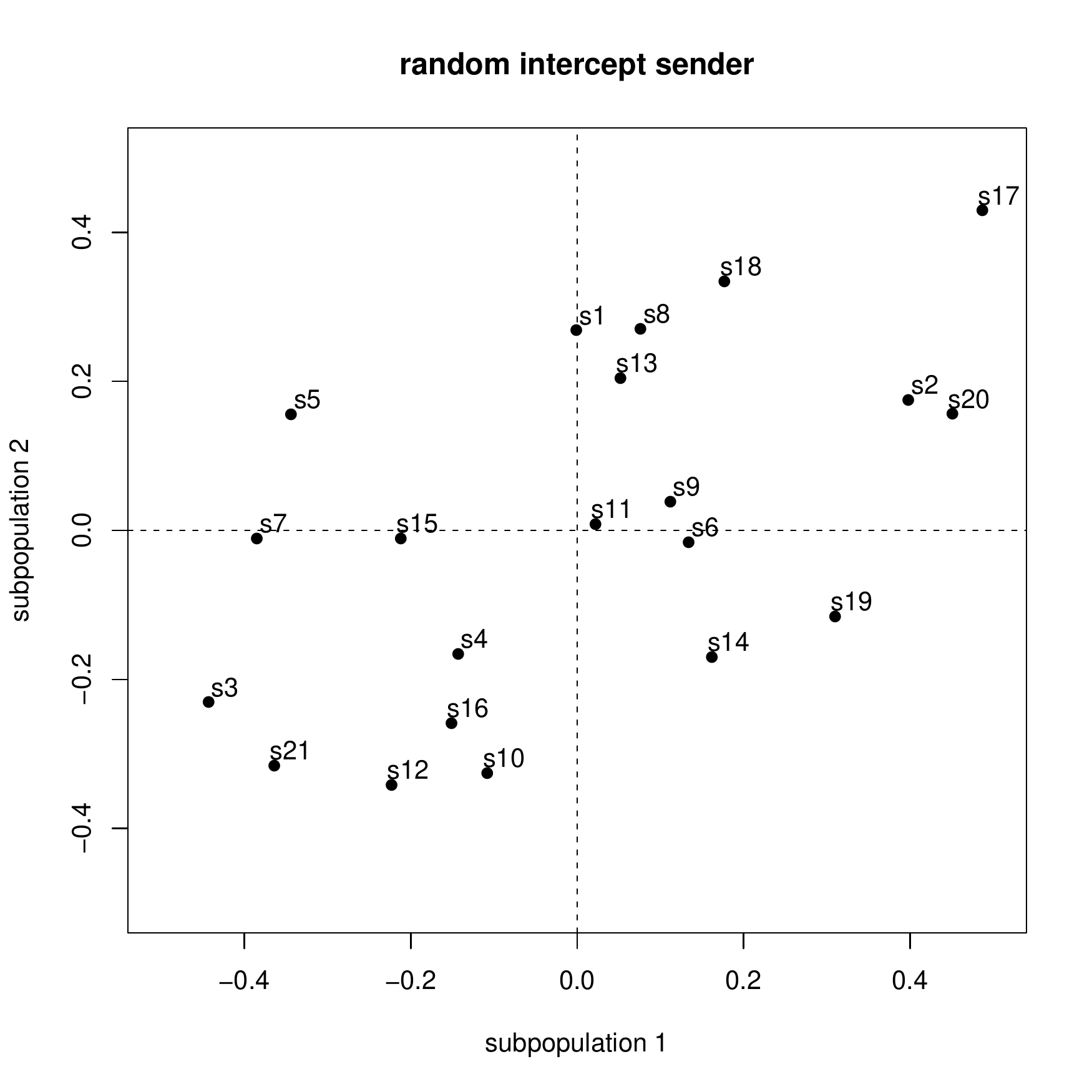} 
\includegraphics[scale=0.425, page = 2]{fig7-appl-ranef.pdf} \end{center}
	\caption{Predicted random intercepts for sender ($\hat{u}_i$) and receiver ($\hat{v}_j$) in subpopulations 1 ($x$ axis) and 2 ($y$ axis).}
\label{fig:appl-ranef}
\end{figure}

\begin{table}[h]
	\centering
\begin{tabular}{|c|p{0.5cm}|p{0.5cm}|p{0.5cm}|p{0.5cm}|}
\multicolumn{5}{c}{Subpopulation 1}\\ \hline
Dept. & \multicolumn{4}{c|}{Dept. receiver}\\
sender & 1 & 2 & 3 & 4\\ \hline
1 & $\bigoplus$ & $\bigominus$ & $\bigominus$ & $-$\\
2 & $\bigominus$ & $\bigoplus$ & $+$ & $\bigominus$\\
3 & $\bigominus$ & $\bigoplus$ & $\bigoplus$ & $+$\\
4 & $-$ & $\bigominus$ & $-$ & $\bigoplus$\\ \hline
\end{tabular}
\hspace{1cm}
\begin{tabular}{|c|p{0.5cm}|p{0.5cm}|p{0.5cm}|p{0.5cm}|}
\multicolumn{5}{c}{Subpopulation 2}\\ \hline
Dept. & \multicolumn{4}{c|}{Dept. receiver}\\
sender & 1 & 2 & 3 & 4\\ \hline
1 & $\bigoplus$ & $\bigominus$ & $\bigominus$ & $-$\\
2 & $\bigominus$ & $\bigoplus$ & $+$ & $\bigominus$\\
3 & $\bigominus$ & $+$ & $\bigoplus$ & $\bigominus$\\
4 & $-$ & $\bigominus$ & $-$ & $\bigoplus$\\ \hline
\end{tabular}
\caption{Sign and significance of the block-interaction parameters $\xi_{rs}^{z_k}$ in cluster 1 (left) and cluster 2 (right). $\bigoplus$ and $\bigominus$ denote parameters significantly different from 0 ($p < 0.05$), $+$ and $-$ parameters with $p > 0.05$. The value and significance of all $\gamma_r^{z_k}$, $\delta_s^{z_k}$ and $\xi_{rs}^{z_k}$ can be found in Table 1 of the Supplementary Materials.}
\label{tab:signs}
\end{table}

Figure \ref{fig:appl-ranef} shows the distribution of the predicted random effects for sender and receiver in model \eqref{form:model-appl}. In the left-hand plot, which displays the sender effect, most points fall in the first and third quadrant; this is an indication that perceivers in the two subpopulations have similar ideas on how many colleagues a certain individual seeks advice from. For example, individual 17 has the highest indegree correction $\hat{u}_i$ in both subpopulations. Similar observations can be made for the right-hand plot; moreover, here we clearly see the different magnitude of the out-degree correction in the two subpopulations, which we already inferred from Table \ref{tab:p2sbm-estimates}.

Table \ref{tab:signs} summarizes the significance and sign of the estimated block-interaction parameters $\xi_{rs}^{z_k}$ in model \eqref{form:model-appl} (more details on $\gamma_r^{z_k}$, $\delta_s^{z_k}$ and $\xi_{rs}^{z_k}$ are provided in Table 1 of the Supplementary Materials). In both clusters we find evidence of a rather strong community structure induced by department affiliation, which results into employees seeking more frequently advice from members of the same department ($\hat{\xi}_{rr}^{m} > 0 \; \forall r \in \{1, 2, 3, 4\} \wedge \forall m \in \{1, 2\}$). All the other block-interactions are typically negative or non-significant, with the exception of advice relationships from department 3 to department 2 in cluster 1 ($\hat{\xi}_{32}^{1} > 0$). Overall, the two subpopulations appear to have a similar, but not identical view of the intensity of advice relationships occurring between members from different departments.

\section{Discussion}

We have developed a model-based clustering approach for populations of networks that specifies a joint statistical model for all graphs in the population and that is capable of identifying subpopulations of graphs which share a similar generative model, but which may still look like quite different networks in edge-space. Building on the fact that GLMs and GLMMs represent a flexible and efficient tool for modelling and estimating a wide variety of generative processes, we have proposed to employ mixtures of GLMs or GLMMs to perform model-based clustering of networks. Estimation of the proposed mixtures of network models can be efficiently carried out by an EM algorithm. The identification of the number of subpopulations that form the mixture has been performed with standard model selection criteria.

Evaluation of the proposed method on simulated data shows that the accuracy of the clustering method strongly depends on the size of the graphs and on the number of clusters, and much less on the number of graphs in the population. In particular, the accuracy increases quickly with the number of vertices and it decreases, as expected, with the number of clusters. The estimation of the number of subpopulations $M$ can be based on the minimization of model selection criteria such as AIC and BIC. As illustrated in Section \ref{sub:sim-inf-crit}, the performance of AIC and BIC appears to be simular when a relatively large number of graphs is available; however, for small $K$ BIC tends to systematically underestimate $M$, so we recommend the use of AIC when dealing with a small number of networks.

The approach presented in this paper is able to consider mixtures of network models that make conditional independence assumptions on the probability of existence of edges. Examples of such models include the $p_1$ model of \cite{holland1981}, the $p_2$ model of \cite{vanduijn2004}, different types of stochastic blockmodels a priori \citep{holland1983,wang1987,signorelli2018}, the loglinear models proposed by \cite{wolfe2013}, the unconstrained model illustrated in Section \ref{sub:unc} and any feasible combination of these models, like the one that we have employed in equation \eqref{form:model-appl}. We note that Exponential Random Graph Models \citep{frank1986} fall outside this class of models as they violate the conditional independence assumption, although quasi-likelihood estimation via a GLM is possible \citep{vanduijn2009}. We have made an attempt to implement this, but the results have been mixed and therefore we do not recommend it in general. 

\section*{Acknowledgments}

We thank two anonymous reviewers, whose useful suggestions and remarks have contributed to improve this paper. The authors gratefully acknowledge funding from the COST Action \emph{European Cooperation for Statistics of Network Data Science} (CA15109).  

\bibliographystyle{apa}
\bibliography{bibliography}

\clearpage
\newpage
\includepdf[pages=1-last]{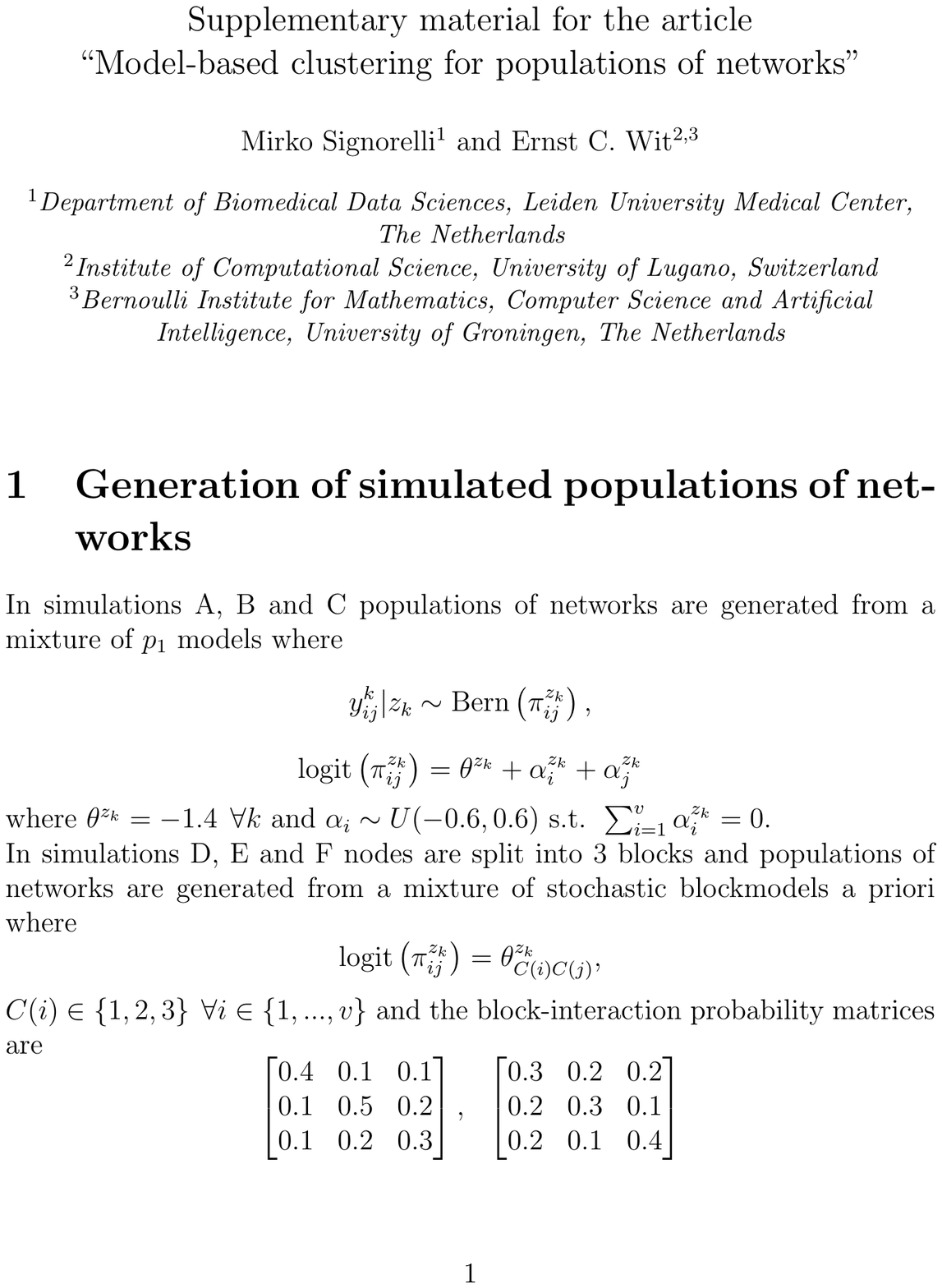}

\end{document}